\documentstyle[12pt]{article}

\newcommand{\be}{\begin{equation}}
\newcommand{\ee}{\end{equation}}

\begin{document}
\pagestyle{empty}
\begin{flushright}
{BROWN-HET-994} \\
{August 1995}
\end{flushright}
\vspace*{5mm}
 \begin{center}
{\bf DECAYING
VACUUM ENERGY AND DEFLATIONARY COSMOLOGY IN OPEN AND CLOSED UNIVERSES}
\\ [10mm]
\renewcommand{\thefootnote}{\alph{footnote}}
J.A.S. Lima$^{1,2,}$\footnote{e-mail:limajas@het.brown.edu}, M.
Trodden$^{1,}$\footnote{e-mail: mtrodden@het.brown.edu.}\footnote{Address from
9/1/95, Center for Theoretical Physics, M.I.T., Cambridge MA. 02139.} \\
[10mm]
\end{center}
\begin{flushleft}
1){\it Physics Department, Brown
University,Providence RI. 02912. USA.} \\
2){\it Departamento de Fisica,
Universidade Federal do Rio Grande do Norte, 59072-970, C.P. 1641 Natal,
Rio Grande do Norte, Brazil.}\\
\end{flushleft}
\begin{center}
{\bf
Abstract}
 \end{center}
\vspace*{3mm}
 We consider a nonsingular
deflationary cosmological model with decaying vacuum energy density in
universes of arbitrary spatial curvature. Irrespective of the value of
$k$, the models are characterized by an arbitrary time scale $H_I^{-1}$
which determines the initial temperature of the universe and the largest
value of the vacuum energy density, the slow decay of which generates
all the presently observed matter-energy of the universe. If $H_I^{-1}$
is of the order of the Planck time, the models begin with the Planck
temperature and the present day value of the cosmological constant
satisfies $\Lambda_I/\Lambda_0 \simeq 10^{118}$ as theoretically
suggested. It is also shown that all models allow a density parameter
$\Omega_0 <2/3$ and that the age of the universe is large enough to
agree with observations even with the high value of $H_0$ suggested by
recent measurements.
\setlength{\textheight}{8.5in}
\newpage\setcounter{page}{1}\pagestyle{plain}
\renewcommand{\thefootnote}{\arabic{footnote}}

\section{Introduction}
A great deal of attention has recently been paid
to cosmological models with a nonvanishing vacuum energy density, or
equivalently a nonzero cosmological $\Lambda$-term. The revival of
interest in these models is physically compelling on both observational
and physical grounds\cite{peebles 84}-\cite{K&T 95}. A large class of
recent observations (the age of the universe, dynamical estimates of the
density parameter, kinematical tests,...etc) consistently point to the
probable existence of an effective vacuum component which, although
incredibly small in comparison with common microscopic scales, is
expected to contribute appreciably to the present large-scale structure
of the universe (for a recent review see\cite{K&T 95}). From a
theoretical standpoint there is also a widespread belief that the early
universe evolved through a cascade of phase transitions, thereby
yielding a present vacuum energy density that is smaller than its value
at Planck times by a factor of at least 118 orders of
magnitude\cite{{Zee 85},{SW 89}}.

On the other hand, since the value of the cosmological ``constant"
$\Lambda_0$ (a subscript~0 denotes the present day value of a quantity)
may be viewed as a remnant of a primordial inflationary stage, it seems
natural to address the following question: Is it possible to describe
the history of the universe accounting for a vacuum energy density that
is high enough to drive inflation at early times and is small enough to
be compatible with observations at late times?

To the best of our knowledge there is no formulation (from first
principles) that provides a satisfactory description of the
time-dependence of $\Lambda$ which presumably occurs as the universe
evolves. In such a situation the classical, phenomenological approach
seems to be a good tool with which to gain some insight into this
question. In fact, models with $\Lambda=\Lambda(t)$ have been the
subject of numerous papers in recent years\cite{O&T 87}-\cite{JM 95}.
Indeed, since the basic motivation is to understand the present day
smallness of the cosmological constant, most scenarios do not attempt to
provide any natural relation between the magnitude of $\Lambda$ at the
beginning of inflation and the present day observational upper bound.

In a previous paper\cite{L&M 94}, we investigated some consequences of a
phenomenological decay law for $\Lambda$ which yielded a partial
solution to the above question. However, since that model was formulated
in the framework of a {\it flat} Friedmann, Robertson-Walker (FRW)
geometry, the results were crucially dependent on that particular
spacetime\cite{singularity}.

In the present paper we wish to demonstrate that the main results of the
previous work remain valid in spacetimes of arbitrary spatial curvature.
To be more precise, there exists a large class of nonsingular
deflationary cosmologies, beginning from the decay of a pure de Sitter
vacuum and subsequently evolving smoothly to a quasi-FRW stage at late
times. The models in this class seem to agree with present cosmological
observations for all values of the curvature parameter $k$. As a general
feature, the process of vacuum decay generates all the matter-radiation
of the present day universe and has the added attraction of
simultaneously solving the same problems that inflation aims to explain.
In addition, as theoretically suggested, the maximum allowed value for
the vacuum energy density is naturally larger than its present value by
about $118$ orders of magnitude.

\section{The Models}
We shall consider metrics described by the general
FRW line element

\be
 ds^2=dt^2-R(t)^2\left(\frac{1}{1-kr^2}dr^2 +r^2d\Sigma^2\right)\ ,
\ee
where $R(t)$ is the scale factor, $d\Sigma^2$ is the area element on
the unit 2-sphere, $k=0,\pm1$ is the curvature parameter and we have
adopted the metric signature convention ($+$,$-$,$-$,$-$). Throughout we
use units such that $c=1$.

In such a background the Einstein field equations (EFE) for the
nonvacuum component plus a cosmological $\Lambda$-term are

\begin{equation} 8\pi G\rho +\Lambda =3\frac{{\dot R}^2}{R^2} +
3\frac{k}{R^2}\ ,
\end{equation}

\begin{equation}
8\pi Gp -\Lambda =-2\frac{{\ddot R}}{R} -\frac{{\dot
R}^2}{R^2}-\frac{k}{R^2}\ ,
\end{equation}
where $\rho$ and $p$ are the
energy density and pressure respectively of the nonvacuum component
which is assumed to obey the $\gamma$-law equation of state

\begin{equation}
p=(\gamma -1)\rho\ , \ \ \ \ \ \ \ \ \ \gamma \in
[1,2]\ .
\end{equation}
As we shall see, regardless of the value of
$k$, a primordial inflationary scenario will automatically be generated
at early times if the vacuum decays according to the following
phenomenological decay ansatz

\be
\rho_V=\frac{\Lambda}{8\pi G} = \beta \rho_T
\left(1+\frac{1-\beta}{\beta}\frac{H}{H_I}\right)\ ,
 \ee
where $\rho_V$
and $\rho_T=\rho_V+\rho$ are the vacuum and total energy densities
respectively, $H\equiv {\dot R}/R$ is the Hubble parameter, $H_I^{-1}$
is the arbitrary time scale of inflation and $\beta$ is a dimensionless
parameter of order unity. For $H=H_I$ equation~(5) reduces to
$\rho_V=\rho_T$ so that we have inflation with no matter-radiation
component ($\rho =0$), while for late times ($H\ll H_I$), $\rho_V
\sim\beta \rho_T$ as is required by recent observations\cite{peebles
84}-\cite{K&T 95}. Since at all times $H\leq H_I$, equation~(5) can be
viewed as the first two terms of a power series expansion of $\rho_V$ in
the parameter $y\equiv H/H_I$. The ansatz~(5) together with
equations~(2) and (3) generalize the model of Freese et. al.\cite{FAF
87} by including the curvature terms and by introducing a time
dependence in the parameter $x\equiv \rho_V/(\rho_V +\rho)$ which here
is given by $x=\beta+(1-\beta)H/H_I$. Of course, at late times $H\ll
H_I$ and this parameter reduces to $x \simeq \beta$ as assumed
in\cite{FAF 87}. Note also that in the flat case $8\pi G\rho_T=3H^2$ and
the flat decaying $\Lambda$-model of Ref.\cite{L&M 94} is readily
recovered, since in this case (5) reduces to (see equation~(1) of
Ref.\cite{L&M 94})

\begin{eqnarray}
\Lambda(H)=3\beta H^2 +3(1-\beta)\frac{H^3}{H_I}\ . \nonumber
\end{eqnarray}

Let us now consider the evolution of the scale factor in these models.
Combining equations (4) and (5) with the EFE we obtain the following
differential equation for $R$ and expression for $\rho$

\be
R{\ddot R} +\Delta({\dot R}^2 + k)\left(1- \frac{(\Delta
+1)}{\Delta}\frac{H}{H_I}\right) = 0\ , \ee \be 8\pi G\rho =
3(1-\beta)\left(H^2 +\frac{k}{R^2}\right)\left(1-\frac{H}{H_I}\right)\ ,
\ee
where

\begin{equation} \Delta \equiv \frac{3\gamma(1-\beta)-2}{2}\ .
\end{equation}

Thus, in the very beginning, where $H=H_I$, (7) gives $\rho=0$ in
accordance with the above qualitative arguments and at late times, where
$H\ll H_I$, the universe is in a quasi-FRW epoch characterized by
$\rho=\rho_T(1-\beta)$ and $\rho_V=\beta \rho_T$ (see equations~(5) and
(7)). Note that $\beta \in [0,1]$ parametrizes the extent to which our
model departs from the standard FRW picture in this phase.

To analyze the solutions of (6) in its various asymptotic regimes it
proves convenient to introduce an {\it effective ``adiabatic index''}

\be
{\tilde \gamma} =\gamma(1-\beta)\left(1-\frac{H}{H_I}\right)\ ,
\ee
so that (6) assumes the general FRW-type form, namely

\be
R{\ddot R} +\left(\frac{3{\tilde \gamma}-2}{2}\right){\dot R}^2 +
\left(\frac{3{\tilde \gamma}-2}{2}\right)k=0\ .
\ee
For $H=H_I$,
equation~(9) gives ${\tilde \gamma}=0$ with (10) reducing to

\begin{equation}
R{\ddot R}-{\dot R}^2 -k =0\ ,
\end{equation}
which
yields the well known de Sitter solutions

\begin{equation}
R(t)=\left\{ \begin{array}{ll}
             H_I^{-1}\cosh(H_I t) & \ \ \ \ \mbox{$k=+1$} \\
             R_* e^{H_I t} & \ \ \ \ \mbox{$k=0$} \ \ ,\\
             H_I^{-1}\sinh(H_I t) & \ \ \ \ \mbox{$k=-1$}
             \end{array}\right.
\end{equation}
Hence, unlike in the
             standard FRW model, the present scenario begins in a pure
             nonsingular de Sitter vacuum with Hubble parameter $H=H_I$.
             Accordingly, equation~(7) gives $\rho=0$ as discussed
             earlier. Note also that in this limit the initial value of
             the $\Lambda$-parameter is $\Lambda_I=3H_I^2$ corresponding
             to a vacuum energy density of $\rho_V=3H_I^2/8\pi G$,
             regardless of the value of $k$. In this way, the initial
             evolution is such that the singularity, flatness and
             horizon problems are simultaneously eliminated.
             Analytically, the ansatz (5) can be viewed as the simplest
             vacuum decay law which destabilizes the initial de Sitter
             configurations given by~(12). As should be expected, no
             dynamic privilege can be associated with a particular
             choice of the curvature parameter of the initial vacuum
             state. All these solutions have constant curvature and are
             unstable in the future. Of course, closed ($k=1$) solutions
             are not of the ``bouncing'' type, rather the universe
             begins its evolution from a closed de Sitter universe.

In the opposite limit, $H\ll H_I$, equation~(9) reduces to ${\tilde
\gamma}=\gamma(1-\beta)$ so that equation~(6) takes the form

\be R{\ddot R} +\Delta {\dot R}^2 +\Delta k =0\ , \ee which is the
general equation for a slightly modified FRW model. There exists a first
integral to this equation, namely

\be {\dot R}^2 = AR^{-2\Delta} - k\ , \ee where the constant $A>0$ in
order that $\rho$ be positive definite in this phase (see equation~(7)).
Parenthetically, such a condition also guarantees the positivity of the
vacuum (and consequently the total) energy density.

Inserting (14) into (5) and (7), the vacuum and the matter energy
density can be expressed for $H\ll H_I$ as

\begin{eqnarray} \rho_V & = & \beta
\rho_{T_0}\left(\frac{R_0}{R}\right)^{3\gamma(1-\beta)}=\beta \rho_T\ ,
\nonumber \\ \rho & = &
(1-\beta)\rho_{T_0}\left(\frac{R_0}{R}\right)^{3\gamma(1-\beta)} \equiv
(1-\beta)\rho_T\ , \end{eqnarray} where $\rho_{T_0}=3A/8\pi G
R_0^{3\gamma(1-\beta)}$. For $\gamma=4/3$ it follows from (15) that the
radiation energy density scales as $\rho_r \sim R^{-4(1-\beta)}$ while
for a dust filled universe ($\gamma =1$) the energy density satisfies
$\rho_d \sim R^{-3(1-\beta)}$. Hence, there is a natural transition from
a vacuum-radiation to a vacuum-dust dominated phase as the universe
expands, just as in the standard FRW model with no-vacuum component. For
the sake of completeness, we remark that in the flat case the evolution
of the scale factor can be analytically described (see Ref.~\cite{L&M
94}, eq.~(10)). In the present notation this is given by

\be H_I t=\ln \left(\frac{R}{R_*}\right) + \frac{2(H_I -H_0)
A^{-1/2}}{3\gamma(1-\beta)} R^{3\gamma(1-\beta)/2}\ .  \ee Hence, in the
very beginning when the logarithm term is dominant, we obtain to a high
degree of approximation $R\simeq R_*e^{H_I t}$ in accordance with our
equation~(12). At late times ($R\gg R_*$ or $H\ll H_I$) one obtains
from~(14) that $A=H_0^2R_0^{3\gamma(1-\beta)}$ with (16) reducing to

\begin{eqnarray} R\sim R_0 \left(3\gamma(1-\beta)\frac{H_0
t}{2}\right)^{2/3\gamma(1-\beta)}\ , \nonumber \end{eqnarray} as
expected (see equation~(15) of Ref.~\cite{L&M 94}). Note also from (5)
and (7) that, irrespective of $k$, both $\rho_V$ and $\rho$ always
satisfy the weak energy condition (e.g. positiveness of the energy
density) during the course of the evolution (see Fig.~1).

It is also worth mentioning that in this scenario there is no
preinflationary stage as in most inflationary variants presented in the
literature\cite{guth 82}-\cite{H&M 82}. In such models the universe
emerges from a radiation dominated FRW-type phase and enters a de Sitter
epoch at a critical temperature due to vacuum domination. In particular,
the existence of such a hot radiation-dominated phase preceding the
vacuum stage means that inflation does not evade the singularity
problem. In connection with this we note that Narlikar and Padmanabhan
proposed a new variant on the ``Creation-field cosmology'' in order to
avoid the singularity problem and other difficulties of the standard
big-bang model\cite{N&P 85}. However, unlike the scenario with vacuum
decay presented here, in such a model the singularity is removed at the
expense of a ``C-field'' of negative energy density which leads to
matter creation.

The initial state of our scenario is the simplest one (constant
curvature) and is physically appealing from a quantum theoretical point
of view. It resembles the early inflationary model proposed by
Starobinskii where the initial de Sitter configurations are supported by
quantum one-loop corrections to the vacuum energy-momentum
tensor\cite{AS 80}. However, unlike the Starobinskii model which evolves
directly from de Sitter to dust domination, the scenario proposed here
contains the same phases of the standard FRW picture and, as we shall
see, has interesting concrete cosmological consequences for the present
vacuum-dust dominated phase (see next section). As a matter of fact,
there have been many suggestions in the literature that the de Sitter
spacetime may be destabilized and decay to ordinary FRW
universes\cite{JRG 82}-\cite{H&T 86}. Of particular interest for us is
the scenario proposed by Gott\cite{JRG 82}. In such a model the universe
begins with the Hawking temperature evolving, at late times, to the
standard FRW model with negative curvature parameter. As we shall see
(see section~4), this connection with the Hawking temperature will be
preserved in our scenario for all values of $k$ since it will define, in
a natural way, the highest values of $\Lambda$ and of the temperature at
the beginning of the universe.

\section{Deflation Confronts Observations} Time varying $\Lambda$ models
usually modify the predictions of the standard FRW picture at both early
and late times, thereby leading to the possibility of constraining the
free parameters of any vacuum decaying universe. In the last section we
saw that the deflationary process driven by the vacuum decay ansatz~(5)
has $H_I$ and $\beta$ as free parameters. However, as we shall see next,
the former does not play any role at late times so that all predictions
of the model concerning the present universe depend only on the
parameter $\beta$.

In order to constrain $\beta$, we shall discuss some dynamical tests.
Following the standard development we define the usual observational
parameters $\Omega_0\equiv8\pi G\rho_0/3H_0^2$ (the matter density
parameter), $q_0 \equiv -R{\ddot R}/{\dot R}^2$ (the decceleration
parameter) and $\Omega_{V_0} \equiv \Lambda_0/3H_0^2$ (the vacuum
density parameter). Using equations~(2), (6) and (7) we obtain the
following expressions for these quantities

\be \Omega_{V_0} = \beta\left(1+\frac{k}{R_0^2H_0^2}\right) + {\cal
O}\left(\frac{H_0}{H_I}\right)\ , \ee

\be \Omega_0 =(1-\beta)\left(1+\frac{k}{R_0^2H_0^2}\right) + {\cal
O}\left(\frac{H_0}{H_I}\right)\ , \ee

\be q_0 = \frac{1-3\beta}{2}\left(1+\frac{k}{R_0^2H_0^2}\right) + {\cal
O}\left(\frac{H_0}{H_I}\right)\ .  \ee As in the flat case (see
equations~(11)-(13) of Ref.~17) the last term on the right hand side of
the above expressions may always be neglected. More precisely, if the
deflationary process begins at the Planck time, $H_I^{-1} \sim
10^{-43}s$ and since $H_0^{-1} \sim 10^{17}s$ it thus follows that
$H_0/H_I \sim 10^{-60}$ while the remaining terms are of order unity.
Even if deflation begins much later, say at $H_I^{-1} \sim 10^{-35}s$ or
$H_I^{-1} \sim 10^{-15}s$ (the respective scales of grand and
electroweak unification in the standard model) one obtains $H_0/H_I \sim
10^{-52}$ and $H_0/H_I \sim 10^{-32}$ respectively. Hence, to a high
degree of accuracy, $H_I$ is unimportant today and equations (17)-(19)
may be written in the simplified forms

\be \Omega_{V_0}=\beta \Omega_{T_0}\ , \ee

\be \Omega_0= (1-\beta)\Omega_{T_0}\ , \ee

\be q_0= \frac{1-3\beta}{2}\Omega_{T_0}\ , \ee where we have introduced
the present day total energy density parameter $\Omega_{T_0}=1+k/R_0^2
H_0^2$. For $\beta=0$ the above expressions reduce to the ones of the
standard FRW model ($\Omega_V=0$), whereas for $\beta \neq 0$ but $k=0$
($\Omega_{T_0}=1$), the results of Ref.\cite{L&M 94} are readily
recovered.

The consistency of the above approximations is easily established by
adding equations~(20) and (21) to obtain
$\Omega_{T_0}=\Omega_0+\Omega_{V_0}$. Further, by eliminating $\beta$
from (21) and (22) it follows that

\be \Omega_0=\frac{2}{3}\Omega_T +\frac{2}{3}q_0\ , \ee which reduces to
the well known result ($\Omega_T=1$) for zero-curvature, (see, for
instance, Ref.~\cite{CLW 92}). As a matter of fact, one can show that
the above relation is quite general, remaining valid for any decaying
$\Lambda$ model. In particular, for $\beta>1/3$ and $\Omega_{T_0} \leq
1$, equations~(22) and (23) imply that flat and open universes satisfy
$\Omega_0 <2/3$, whereas for closed models this holds only if the
additional constraint $1<\Omega_{T_0}<2/3(1-\beta)$ is imposed. Note
also that (21) can be rewritten as

\be \frac{k}{R_0^2}=\left(\frac{\Omega_0}{1-\beta}-1\right)H_0^2\ , \ee
explaining how the low-energy problem is alleviated in such a scenario,
since this is the same as the usual FRW expression but with an effective
matter density parameter $\Omega_{eff}=\Omega_0/(1-\beta)$. As we show
below, this fact allows us to easily solve the age problem in this
context.

The most physically appealing observational data calling for the
investigation of cosmological ``constant'' models involves the so-called
``{\it age problem}''. In short, the ages of the oldest globular
clusters are estimated to be $16\pm 3\ $Gyr while, paradoxically, a
large value of the Hubble parameter (the natural inverse time scale of
the FRW geometries) centered at $H_0=80\pm 17\ kms^{-1}Mpc^{-1}$ is
favored by recent measurements\cite{WLF 94}. The root of the
conflict is that in the standard flat FRW model this value of $H_0$
corresponds to an expansion age ($t_0=2/3H_0$) of nearly $8.3\ $Gyr.
The situation is even worse if the data of Pierce et.al.\cite{PWM 94}
($H_0=87 \pm 7 kms^{-1}Mpc^{-1}$) are considered. In this case the age is
only $7.3$Gyr.

Such a paradox is easily resolved in the present decaying
$\Lambda$-model. As in the flat case\cite{L&M 94}, the time required by
the deflationary process is much longer than the corresponding quasi-FRW
phase. Note that, even in the open case, the spacetime is regular at the
horizon ($t=0$) and can be continued beyond this point\cite{AS 80}.
Computing the value of the constant $A$ in terms of the observational
parameters (see equation~(14)), it is straightforward to conclude that a
lower bound for the age of the universe is given by

\be
t_0 = H_0^{-1}\int_{x_{min}}^1\frac{dx}{\sqrt{1-\frac{\Omega_0}{1-\beta}
+\frac{\Omega_0}{1-\beta}x^{-(1-3\beta)}}}\ ,
\ee
where $x_{min}$ is the smallest value of $x$ for which the integrand remains
real.
In particular, for flat
models ($\Omega_{T_0}=1$, $\Omega_0=1-\beta$, $x_{min}=0$) this
expression yields

\be
t_0 = \frac{2}{3(1-\beta)} H_0^{-1}\ ,
\ee
in agreement with
Ref.~\cite{L&M 94}. In what follows all estimates will be made using the
somewhat more conservative data of Friedman et.al.\cite{WLF 94}. Figure~2
shows the age of the universe (in units of
$H_0^{-1}$) as a function of $\Omega_0$ for some selected values of
$\beta$. The above mentioned observations restrict the dimensionless age
parameter $H_0t_0$ (which is 2/3 in the standard flat FRW model) to the
interval

\be
0.85 \leq H_0t_0 \leq 1.91\ ,
\ee
which should be compared with the
rather conservative bounds ($0.6 \leq H_0t_0\leq 1.4$) adopted in
Ref.~\cite{L&M 94}. From (26) and (27) it is easily seen that
deflationary models solve the age conflict if the allowed values of
$\beta$ are constrained to be $0.21\leq \beta \leq 0.64$. It is
interesting that for $\beta$ in this range the values of our
observational parameters are restricted to satisfy (see equations~(20)-(22))

\be
0.63H_0^2 \leq \Lambda_0 \leq 1.92 H_0^2\ ,
\ee

\be
0.36 \leq \Omega_0 \leq 0.79\ ,
\ee

\be
-0.46 \leq q_0 \leq 0.18\ ,
\ee
It is worth noting that not only is
$\Lambda_0$ below the presently accepted upper bound ($\Omega_V \leq
0.8$, $\Lambda_0 \leq 2.4 H_0^2$), but the low-energy problem becomes
much less serious. As a matter of fact, if the ``best-fit'' model
consists of $\Omega_T=1$ with $\Omega_{V_0}=0.7\pm 0.1$ and
$\Omega_0=0.3\pm 0.1$, as claimed by some authors
\cite{{peebles 84},{K&T 95}}, then
$\beta=0.8\pm 0.1$ and from (25) the age
problem is more easily resolved.

As is well known, vacuum decay $\Lambda$-models predict both matter and
entropy production\cite{O&T 87}-\cite{L&M 94}. The present day rate of
the former is readily obtained from the energy conservation law
${T^{\mu\nu}}_{;\nu}=0$ expressed as

\be {\dot \rho} + 3H(\rho+p)=-\frac{1}{8\pi G}{\dot \Lambda}\ , \ee or
equivalently, from (4)

\be \frac{1}{R^{3\gamma}}\frac{d}{dt}(\rho R^{3\gamma}) = -\frac{1}{8\pi
G}{\dot \Lambda}\ .  \ee

At the present time ($H \ll H_I$, $\gamma =1$), the matter production
rate is easily computed. Combining equations~(5) and (14) it follows
that

\begin{eqnarray} {\dot \Lambda}(t_0)= -9(1-\beta)\beta H_0\left(H_0^2
+\frac{k}{R_0^2}\right) +{\cal O}\left(\frac{H}{H_I}\right)\ , \nonumber
\end{eqnarray} (in Ref.~\cite{L&M 94} the factor $\beta$ is absent) and
using (21) we have

\be \frac{1}{R_0^3}\left.\frac{d}{dt}(\rho R^3)\right|_{t_0} =3\beta H_0
\rho_0\ , \ee as previously obtained (see equation~(17) of
Ref.~\cite{L&M 94}). Therefore the present matter creation rate does not
depend explicitly on the curvature parameter. Observe that the factor
$3\rho_0 H_0 \sim 10^{-41}\ gcm^{-3}yr^{-1}$ is merely the creation rate
appearing in the steady state model and thus lies far below detectable
limits. Note also that (31) may be rewritten to yield an expression for
the rate of entropy production in this model\cite{{O&T 87},{A-R 92}} as

\begin{eqnarray} T\frac{dS}{dt}=-\frac{{\dot \Lambda}R^3}{8\pi G}\ .
\nonumber \end{eqnarray} In particular, for $H=H_I$ we have ${\dot S}=0$
and at late times ($H\ll H_I$) it is easy to see that

\begin{eqnarray} \frac{dS}{dt}=\frac{3\beta H_0 \rho_0 R_0^3}{T_0}\ .
\nonumber \end{eqnarray}

At this point it is appropriate to make a remark concerning baryogenesis
in these models. The important observational quantity for baryogenesis
is the baryon to entropy ratio $\eta \equiv n_b/s$ where $n_b$ is the
excess number density of baryons over antibaryons and $s$ is the entropy
density. Since in our models both the temperature-scale factor
relationship and the entropy density at a given temperature differ from
those in the standard FRW picture we expect there to be implications for
all baryogenesis scenarios. Naturally, similar remarks can also be made
concerning the predictions of light element abundances from primordial
nucleosynthesis. In this context we note that the results of Freese et
al.\cite{FAF 87} indicate very tight bounds on the parameter $\beta$,
thereby leading to the conclusion that the universe cannot be vacuum
dominated for times later than about $t\sim 1s$. However, such a result
is in conflict with a wealth of observational indications of a vacuum
component in the presently observed universe\cite{K&T 95}. This issue
will be addressed elsewhere.

\section{Final Comments} The study of cosmological models with decaying
vacuum energy density has at least a twofold motivation: to determine
how the high value of the vacuum energy density that drove inflation
became so small at present and to solve the age problem which, by the
latest measurements, plagues the standard model for all values of the
curvature parameter.

In this paper, the FRW-flat cosmological scenario driven by decaying
vacuum energy density as proposed in Ref.~\cite{L&M 94} has been
extended to include the curvature terms. Our deflationary model provides
an interesting cosmological history that evolves in three stages: First,
an unstable de Sitter configuration is supported by the largest values
of the vacuum energy density $\rho_V=3H_I^2/8\pi G$. Initially, for all
values of $k$, there is no matter or radiation in the usual sense. This
happens because $H_I$ is the maximum allowed value for the Hubble
parameter and at $H=H_I$ the model yields $\rho=0$ (see equation~(7) and
Fig.~1). As we shall see in a moment, this de Sitter initial state is an
indispensable ingredient in harmonizing the scenario with the so-called
``{\it cosmological constant problem}''. Secondly, the de Sitter
configuration evolves to a quasi-FRW vacuum-radiation dominated phase,
thereby naturally solving the horizon and other well-known problems in
the same manner as in inflation. This is achieved simply by taking
$\gamma =4/3$ in all equations at early times. There genuinely is no
flatness problem in this scenario. Such a problem appears in the
standard FRW model because the total entropy ($S\sim T^3R^3$) is
constant with $T\propto t^{-1/2}$ and $R\propto t^{1/2}$ at times of
order the Planck time\cite{B&G 87}. As we have shown, these conditions
are not satisfied in our model. The burst of entropy and matter is
provided by the decay of the vacuum which is solely responsible by the
initial de Sitter configurations for $k=0,\pm 1$. The status of the FRW
class of geometries is recovered in the sense that only observations can
decide if the universe is flat, negatively curved or positively curved
nowadays. In other words, the flat ($k=0$) geometry is no longer
theoretically favored. Such an evolution, which for $k=0$ is exactly
described by equation~(16), can also be viewed as a noteworthy solution
to the ``graceful exit'' problem of old inflation\cite{barrow}. Finally,
the transition from the vacuum-radiation to the vacuum-dust stage occurs
in the same manner as in the standard cosmology.

The ansatz~(5) can mathematically be considered as the simplest
$\Lambda(t)$ which destabilizes the initial de Sitter configurations. As
is well known, in the spirit of quantum cosmology it seems natural to
expect negligibly small deviations from such a highly symmetric
spacetime at the beginning of the universe (see \cite{G 93} and
references therein). In connection with this we recall that quantum
effects in the de Sitter spacetime give rise to a geometrothermodynamic
equilibrium state characterized by the Gibbons-Hawking temperature $k_B
T=\hbar(\Lambda/12\pi^2)^{1/2}\ $\cite{G&H 77}. In the present case
$\Lambda_I=3H_I^2$ so that the initial temperature of our scenario is
given by

\be
T_I=\frac{\hbar H_I}{2\pi k_B}\ ,
\ee
where $H_I^{-1}$, the
arbitrary time scale of the de Sitter state is not fixed by the model.
This allows us to make the natural choice that $H_I^{-1}$ be of the
order of the Planck time. Indeed, in the framework of quantum cosmology,
many authors have suggested that the spontaneous birth of the universe
leads naturally to a de Sitter stage with $H^{-1}\sim t_p$ or
equivalently $\rho_V=\rho_{\rm PLANCK}$ (see for example \cite{AV 82}).
It is remarkable that such a choice, say $H_I=2\pi t_p^{-1}$, has two
interesting consequences: First, from (34) the initial temperature of
the universe is just the Planck temperature

\begin{eqnarray}
T_I=\frac{1}{k_B}\sqrt{\frac{\hbar}{G}}\ . \nonumber
\end{eqnarray}
Further, since our model essentially predicts
$\Lambda_I/\Lambda_0\sim (H_I/H_0)^2$ we obtain $\Lambda_0\sim
10^{-118}\Lambda_I$ as theoretically expected. This generalizes the
results of Ref.~\cite{L&M 94} for all values of the curvature parameter.
The vacuum energy density decays from $\rho_V=\rho_{\rm PLANCK}$ to the
present value $\rho_V \simeq \beta H_0^2$, thereby generating all the
matter-energy filling the observable universe. Presumably, the specific
form of the constants $H_I$ and $\beta$ will be furnished by a
fundamental particle physics model of decaying vacuum energy density.

\newpage
\vspace{1cm}
\begin{center}
\bf Acknowledgements
\end{center}
\vspace{5mm}
It is a pleasure to thank Robert Brandenberger and Jackson Maia for many
valuable suggestions and a critical reading of the manuscript. Many
thanks are also due to Raul Abramo, Richhild Moessner and Andrew
Sornborger for their permanent stimulus and interest in this work. One
of us (JASL) is grateful for the hospitality of the Physics Department
at Brown University. This work was partially supported by the Conselho
Nacional de Derenvolviments Cientifico e Tecnologico - CNPq (Brazilian
Research Agency), and by the US Department of Energy under grant
DE-F602-91ER40688, Task A.

\newpage
\begin{center}
\bf Figure Captions
\end{center}
Figure 1: The vacuum (full line) and matter (dashed line) energy densities
as a function of the Hubble parameter in units of $H_I$. Note that in these
units the present value, $H_0$, is essentially zero.
\vspace{5mm}
\newline
Figure 2: The age of the universe in units of $H_0^{-1}$ as a function of
$\Omega_0$ for selected values of $\beta$. The two horizontal lines on the
plot are the allowed range of the age from observations (see equation~27).
Note that for $0.21 \leq \beta \leq 0.64$ the age problem is
solved.
\end{document}